\documentclass[prl,aps,twocolumn,floats,nofootinbib]{revtex4-1}
\usepackage{amsmath,amssymb,graphicx,psfrag}
\usepackage[colorlinks,linkcolor={blue},citecolor={blue},urlcolor={blue}]{hyperref}

\begin{document}
\renewcommand{\ni}{{\noindent}}
\newcommand{\dprime}{{\prime\prime}}
\newcommand{\be}{\begin{equation}}
\newcommand{\ee}{\end{equation}}
\newcommand{\bea}{\begin{eqnarray}} 
\newcommand{\eea}{\end{eqnarray}}
\newcommand{\la}{\langle}
\newcommand{\ra}{\rangle} 
\newcommand{\dg}{\dagger}
\newcommand\lbs{\left[}
\newcommand\rbs{\right]}
\newcommand\lbr{\left(}
\newcommand\rbr{\right)}
\newcommand\f{\frac}
\newcommand\e{\epsilon}
\newcommand\ua{\uparrow}
\newcommand\da{\downarrow}

\newcommand{\Eqn}[1] {Eqn.~(\ref{#1})}
\newcommand{\Fig}[1]{Fig.~\ref{#1}}

\title{Many-body localization and enhanced  non-ergodic sub-diffusive regime in the presence of random long-range interactions}
\author{Yogeshwar Prasad and Arti Garg}
\affiliation{Condensed Matter Physics Division, Saha Institute of Nuclear Physics, 1/AF Bidhannagar, Kolkata 700 064, India}
\vspace{0.2cm}
\begin{abstract}
\vspace{0.3cm}
	{We study many-body localization (MBL) in a one-dimensional system of spinless fermions with a deterministic aperiodic potential in the presence of random interactions $V_{ij}$ decaying as power-law $V_{ij}/(r_{ij})^\alpha$ with distance $r_{ij}$. We demonstrate that MBL survives even for $\alpha <1$ and is preceded by a broad non-ergodic sub-diffusive phase. Starting from parameters at which the short-range interacting system shows infinite temperature MBL phase, turning on random power-law interactions results in many-body mobility edges in the spectrum with a larger fraction of ergodic delocalized states for smaller values of $\alpha$. Hence, the critical disorder $h_c^r$, at which ergodic to non-ergodic transition takes place increases with the range of interactions. Time evolution of the density imbalance $I(t)$, which has power-law decay $I(t) \sim t^{-\gamma}$ in the intermediate to large time regime, shows that the critical disorder $h_{c}^I$, above which the system becomes diffusion-less (with $\gamma \sim 0$) and transits into the MBL phase is much larger than $h_c^r$. {\it In between $h_{c}^r$ and $h_{c}^I$ there is a broad non-ergodic sub-diffusive phase, which is characterized by the Poissonian statistics for the level spacing ratio, multifractal eigenfunctions and a non zero dynamical exponent $\gamma \ll 1/2$}. The system continues to be sub-diffusive even on the ergodic side ($h < h_c^r$) of the MBL transition, where the eigenstates near the mobility edges are multifractal. For $h < h_{0}<h_c^r$, the system is super-diffusive with $\gamma >1/2$. The rich phase diagram obtained here is unique to random nature of long-range interactions. We explain this in terms of the enhanced correlations among local energies of the effective Anderson model induced by random power-law interactions.}
         
\vspace{0.cm}
\end{abstract} 
\maketitle
\section{I. Introduction}

Many-body localization (MBL) has been a topic of immense interest in condensed matter physics. Though generic interacting clean systems are diffusive, the absence of diffusion is a hallmark of systems that undergo Anderson localization~\cite{Anderson} and many-body localization~\cite{Basko,Gornyi,Huse_rev,Abanin_rev,Abanin2,Alet_rev,Ehud_rev}. The MBL phase is a non-ergodic phase in which local observables do not thermalize leading to violation of eigenstate thermalisation hypothesis (ETH)~\cite{Deutsch,Srednicki,Rigol}. MBL phase has been shown to have similarity with integrable systems with an extensive number of local integrals of motion~\cite{Abanin,Mueller}.

In most of the MBL systems, there occurs a transition from an ergodic delocalized phase to the non-ergodic MBL phase as the disorder strength increases compared to the interaction strength~\cite{Alet_rev,Imbrie,Huse2007,Huse2010,Huse2013,Bardarson,Bera,Alet,Pal,Subroto,shastry,ME1,garg,Bera2,sdsarma2019,expt,expt2}; however it is also possible to have an intermediate non-ergodic extended phase. Such a phase has been described in numerical studies on Josephson junction arrays~\cite{JJA}, quantum random energy model~\cite{QREM} and in a one dimensional fermionic system with spin-orbit couplings~\cite{SO}. In conventional models of MBL of spin-1/2 particles or spinless fermions on a chain, the non-ergodic extended phase has been realized~\cite{Serbyn,Santos} for very narrow range of parameters raising questions about its stability in the thermodynamic limit; though in models with quasi-periodic potentials which have single particle mobility edges, it has been argued that the non-ergodic extended phase survives in the thermodynamic limit~\cite{Sdsarma,Subroto-NEE}.  Here we present a novel route to realize a broad non-ergodic subdiffusive phase in the presence of random long-range interactions in a system where all the single particle states are localized.
\begin{figure}[h!]
  \begin{center}
    \hspace{-1cm}
  \includegraphics[width=3.65in,angle=0]{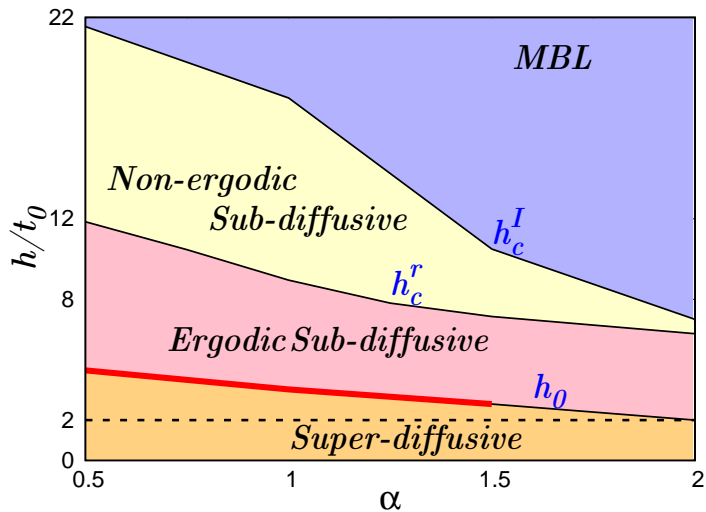}
  \caption{Phase diagram of the model in \Eqn{eqn:model} in $\alpha-h$ plane. For very weak disorder strength $h <2t_0$, the system is ergodic (level spacing ratio obeys Wigner-Dyson statistics)  and super-diffusive (the density imbalance $I(t)\sim t^{-\gamma}$ with $\gamma >1/2$) for all ranges of interactions. For long range interactions ($\alpha < 2$) the super-diffusive phase extends up to $h_0>2t_0$. For $h>h_0$, the system shows sub-diffusive transport with $\gamma <1/2$ though a narrow diffusive regime ($\gamma =1/2$) is expected close to $h_0$ for $\alpha <2$ shown as red shaded region in the phase diagram. As the disorder strength is increased further, ergodic to non-ergodic transition takes place at $h_{c}^r$, which is determined from the level spacing statistics. $h_{c}^r$ increases monotonically as $\alpha$ decreases. The non-ergodic side of the transition is not MBL as indicated by slow dynamics in the quantum quench, rather it is sub-diffusive with multifractal eigenstates. A much stronger strength of disorder $h_{c}^I$ is required for transition into the MBL phase where $\gamma \sim 0$, with $h_{c}^I$ increasing as $\alpha$ decreases.}
\label{pd}
\end{center}
\end{figure}

Effect of random interactions on MBL is perplexing and interesting. On one side, random interactions in the absence of any random one-body potential (e.g. random magnetic field for spin systems) have been shown to cause MBL beyond the picture of local integrals of motion for the case of nearest neighbour~\cite{Sarma_random} as well as long-range interactions~\cite{garg_SG}. On the contrary, sufficiently long-ranged random interactions in the presence of random magnetic field have been argued to cause delocalization for any infinitesimal strength of interactions~\cite{Burin,Dipole,xy,Mirlin,Yevgeny2020}. For example, in the Heisenberg and XXZ model (with long-range zz term) with a random magnetic field and random-interactions $J_{ij} = \pm J$ decaying as a power-law $J_{ij}\sim r_{ij}^{-\alpha}$ with distance, MBL phase does not survive for $\alpha < 2d$~\cite{Burin,Mirlin,Yevgeny2020} where $d$ is the dimension of the system. We emphasize that the absence of MBL in these models is not linked only to the long-range nature of interactions, but the fact that coefficients of interactions are random plays a crucial role in delocalizing the system, especially for the case of $zz$ interactions. This is evident from the fact that in the one-dimensional XXZ and Ising model with uniform coefficients of the power-law $zz$ term MBL has been shown to exist for $\alpha <2$ both theoretically~\cite{Dipole,Sarma2015,Heyl,logan_lr} and experimentally~\cite{tc1,tc,ions2}. Similarly, in system of spin-less fermions in the presence of power-law interactions with uniform coefficients and nearest neighbour hopping, MBL exists even for $\alpha <1$~\cite{garg_lr,note}. In this work, we study the effect of random power-law interactions on a one-dimensional model of spinless fermions in the presence of an aperiodic potential and show that transition from the ergodic extended phase to the MBL phase occurs via an intervening non-ergodic extended phase, as the strength of aperiodic potential is increased. The intermediate non-ergodic phase is characterized with multifractal states and subdiffusive transport and its width increases with the range of interactions.

To be specific, in this work we study a half-filled model of spinless fermions in one dimension in the presence of an aperiodic deterministic potential~\cite{Fishman,Sarma1990,Sarma_nonint,garg,garg_lr} and random power-law interactions $V_{ij}/r_{ij}^\alpha$ among fermions. Interestingly, power-law interactions with random coefficients significantly modify the correlations among local energies of the effective Anderson model in the Fock space such that a much stronger aperiodic potential is required to achieve non-ergodic phase, characterised by the Possonian statistics for level spacing ratio, for systems with longer range interactions. Furthermore, a broad regime of this non-ergodic phase continues to have subdiffusive transport before the system becomes diffusion-less in the MBL phase. The main findings of our work, that are presented in the phase diagram of \Fig{pd}, are summarized below. (1) Starting from the parameters at which the system in the presence of nearest neighbour interaction shows infinite temperature MBL phase, turning on power-law interactions with random coefficients results in the formation of the many-body mobility edges separating the non-ergodic localized states from the ergodic extended states. The fraction of ergodic extended states is larger for the longer range interactions (i.e. smaller values of $\alpha$) such that the critical disorder, $h_{c}^r$, at which the level spacing ratio for the entire spectrum obeys Poissonian statistics increases with increase in the range of interactions. (2) The dynamics of the system after a quench starting from a charge density wave ordered initial state demonstrates that a part of the ergodic phase below $h_c^r$ is sub-diffusive where the density imbalance $I(t) \sim t^{-\gamma}$ with $\gamma <1/2$ though for $h<h_0<h_c^r$ the system is super-diffusive with $\gamma >1/2$. For short range interactions ($\alpha >2$), $h_0$ coincides with the localization transition point of the non-interacting system  but with increase in the range of interactions $h_0$ increases. (3) For $h > h_c^r$ there exists a broad non-ergodic sub-diffusive phase, intervening the ergodic sub-diffusive phase and the MBL phase, which is characterized by Posissonian statistics for level spacing ratio and exponent $\gamma \ll 1/2$ and whose width increases with decrease in $\alpha$. (4) Analysis of eigen-function statistics shows that for $ h_c^r < h < h_c^I$, where the system is in non-ergodic sub-diffusive phase, eigenstates in the middle of the spectrum are multifractal. (5) Eventually at $h_c^I$, the system enters into the MBL phase which has non-ergodic localized states and is characterized by the absence of power-law decay in the imbalance (that is $\gamma=0$).

The rest of the paper is organized as follows. In Section II, we introduce the model explored in this work. In section III, we analyze correlation among local energies of the effective Anderson model as a function of the range of random power-law interactions. In section IV we discuss results for the level spacing statistics and identify the ergodic to non-ergodic transition point $h_c^r$ which is a function of the range of interaction. In section V, we describe dynamics of the system after a quench starting from a charge density wave ordered initial state both below the $h_c^r$ as well as above it. Section VI presents results for the eigen-function statistics based upon the analysis of inverse participation ratio (IPR). Finally we summarize our results and conclude with some remarks and open questions.

 \section{II. Model}
 \label{sec:model}
We study a model of spin-less fermions in one-dimension described by the following Hamiltonian  
\bea
H=-t_0\sum_{i}[c^\dagger_ic_{i+1}+h.c.] + \sum_i h_i n_i \nonumber \\
+\sum_{j>i} V_{ij} \frac{n_in_{j}}{|r_i-r_j|^\alpha}
\label{eqn:model}
\eea
Here $t_0$ is the nearest neighbor hopping amplitude  with open boundary conditions, and $h_i$ is the on-site potential of the form $h_i=h \cos(2\pi\beta i^n+\phi)$ where $\beta = \frac{\sqrt{5}-1}{2}$ is an irrational number and $\phi$ is an offset~\cite{Fishman,Sarma1990,Sarma_nonint}. $V_{ij}$ is coefficient of the power-law interaction term between fermions, chosen randomly from a uniform distribution $[-V,V]$ where $V$ is measured in units of $t_0$. In the non-interacting limit, $V_{ij}=0$, for $n=1$, this model maps to the well known Aubry-Andre (AA) model~\cite{AA}, where all the single particle states are delocalized (localised) for $h < 2t_0$ ($h> 2t_0$). For $n < 1$, the system has single particle mobility edges at $E_c=\pm|2t_0-h|$ for $h < 2t_0$~\cite{Sarma1990}. For $h>2t_0$, all the single particle states are localized for any value of $n$. In this work, we choose to work with this model rather than fully random disorder because there are no rare-region effects due to the deterministic aperiodic potential, which are generally considered to be the cause for sub-diffusive transport. This will help in understanding the role of random long-range interactions on MBL more clearly. Also by tuning parameters, one can study system with single particle mobility edges or without it. In this work we chose to work with  $n=0.5$.

The model in \Eqn{eqn:model} with nearest-neighbour repulsion between fermions (i.e. $V_{ij}=V$ for $j=i+1$ and zero otherwise) has been studied earlier~\cite{Subroto,garg}. For $h < 2t_0$, where the non-interacting system has single particle mobility edges, the interacting system with nearest-neighbour (NN) interactions shows MBL only if the chemical potential does not lie between the two single particle mobility edges, that is for special dopings away from half-filling. Though for $h>2t_0$, the NN interacting system can show MBL at any filling for weak interactions. For very strong disorder $h\gg 2t_0$, the NN interacting system shows an infinite temperature MBL phase where all the many-body states are localized. In this work, we focus on half filled limit of the model in \Eqn{eqn:model} to explore the competing effects of random interactions and the aperiodic potential. In order to understand the physics of random power-law interactions, we map this model to an effective Anderson model in the many-body Fock space.

\begin{figure}[h!]
  \begin{center}
    \vskip0.5cm
\includegraphics[width=3.1in,angle=0]{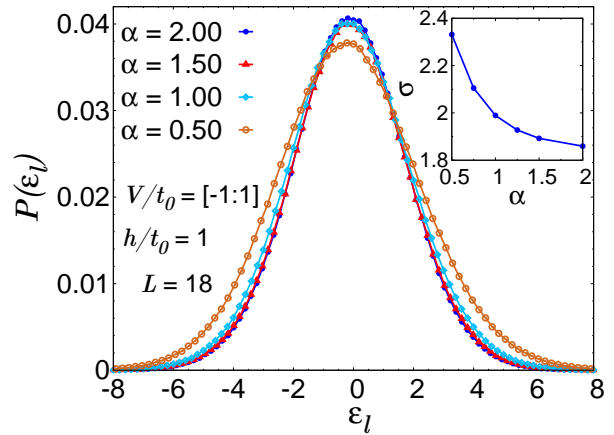}
\caption{Probability distribution of the local energy, $\epsilon_l$, of the effective Anderson model in \Eqn{eqn:heff}. The width of the distribution increases as $\alpha$ decreases. Inset shows the standard deviation $\sigma$ as a function of $\alpha$.}
\label{local_en}
\vskip-1cm
\end{center}
\end{figure}
\section{III. Correlation among local energies of the effective Anderson model in the many-body Fock space}
We map the Hamiltonian in \Eqn{eqn:model} to an effective Anderson model defined in the Fock space of spinless fermions which has $^LC_{L/2}$ configurations for a half-filled chain of $L$ sites. The effective Hamiltonian in the Fock space basis has the following form:
\be
H_{eff} = \sum_l \epsilon_l |l\ra \la l|+ \sum_{lm} \hat{T}_{lm}|l\ra \la m|
\label{eqn:heff}
\ee 
with $\epsilon_l = \sum_i h_i \la l|n_i|l \ra +\sum_{j>i} V_{ij}\f{\la l|n_in_j|l \ra}{(j-i)^\alpha}$ and $\hat{T}_{lm} = -t_0\sum_i \la l|c^\dagger_ic_{i+1}+h.c.|m $. Here ${|l\ra}$ represents configurations in the Fock space which are specified by the occupancies at each site $|{n_i}\rangle$ where $n_i$ is 1 or 0 if the site $i$ in the real space is occupied  or unoccupied.
Let us analyze the local energy, $\epsilon_l$, that has contribution from interactions among all the particles in the system. \Fig{local_en} shows the probability distribution $P(\epsilon_l)$ of the local energy for $h=t_0$ and $V_{ij} \in [-1,1]$ for various values of $\alpha$. As $\alpha$ decreases, the width of the distribution increases and hence the standard deviation, $\sigma$, increases as shown in the inset. This implies that as the range of random power-law interactions increases, the strength of effective disorder in the Anderson model (\Eqn{eqn:heff}) on Fock space also increases.
\begin{figure}[h!]
\begin{center}
\includegraphics[width=3.3in,angle=0]{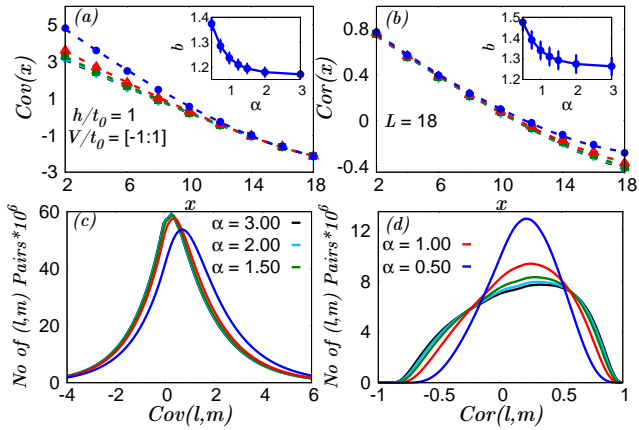}
\caption{Bottom panels show the distribution of covariance $Cov(l,m)$ and the correlation $Cor(l,m)$ among local energies of the effective Anderson model for $|l\ra,|m\ra$ configurations of the Fock space. As $\alpha$ decreases, $Cov(l,m)$ for most of the pairs increases though the number of Fock state pairs having $-0.3<Cor(l,m)<0.6$ increases significantly. Top panels show the plot of $Cov(x)$ and $Cor(x)$ as a function of the Hamming distance $x$ between a pair of Fock states $(|l\ra,|m\ra)$. Note that both $Cov(x)$ and $Cor(x)$ $\sim (1-x/L)^b$ with $b \sim 1$ for the short range model and it increases as the range of interaction increases. The data shown is for $h=t_0$ and $V_{ij}\in [-1,1]$ for $L=18$.}
\label{corr}
\vskip-1cm
\end{center}
\end{figure}
Thus, naively one would expect enhanced localization in a system with random longer range interactions. In fact in a recent study it has been shown that random nearest-neighbour interactions between one-dimensional spin-less fermions alone, without any onsite-disorder, can stabilize MBL~\cite{Sarma_random}.  
But random-power-law interactions also affect the correlation of the local energies $\epsilon_l$. Correlation among local energies have been shown to be crucial for MBL~\cite{Subroto_corr,Logan_corr}, and hence it is important to understand how these correlations are modified in the presence of random power-law interactions. 

The covariance between a pair of Fock space onsite energies is defined as $Cov(l,m) = \la \epsilon_l\epsilon_m \ra - \la\epsilon_l\ra \la\epsilon_m\ra$ where $\la \ra$ is the average over various independent disorder configurations of the aperiodic potential as well as the random power-law interactions. The left bottom panel of \Fig{corr} shows the number of pairs of the Fock states having a certain value of $Cov(l,m)$. As the range of interaction increases, most probable value of the covariance also increases. The top left panel of \Fig{corr} shows covariance as a function of the Hamming distance $x$ between a pair of configuration $|l\ra$ and $|m\ra$. $Cov(x) \sim a(1-x/L)^b$, with the exponent $b \rightarrow 1$ as the range of interaction decreases which is consistent with earlier studies on short range interacting systems. Also $Cov(x)$ for $2 < x< L/2$ increases as $\alpha$ decreases. 
\begin{figure}[h!]
  \begin{center}
\includegraphics[width=3.3in,angle=0]{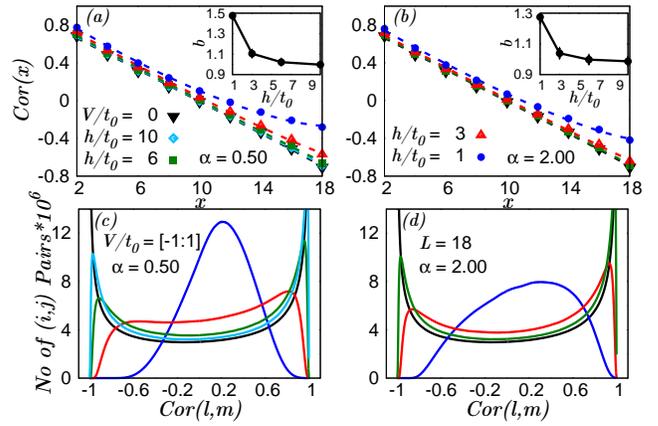}
\caption{Bottom panels show the distribution of correlation $Cor(l,m)$ among local energies of the effective Anderson model for various values of the disorder $h$ and $V_{ij}\in [-1,1]$. The left panel shows results for $\alpha=0.5$ while the right panel shows results for $\alpha=2$. As $h$ increases, probability for having smaller values of $|Cor(l,m)|$ decreases though the larger values of $|Cor(l,m)|$ become more probable. As $h$ is increased further, distribution of $Cor(l,m)$ eventually approaches that for the non-interacting $V=0$ case. Top panels show $Cor(x)$ as a function of the Hamming distance $x$. $Cor(x) \sim (1-x/L)^b$ and $b \rightarrow 1$ as $h$ increases.}
\label{corr2}
\vskip-1cm
\end{center}
\end{figure}
A related quantity useful to study is the correlation among local energies $Cor(l,m) = Cov(l,m)/(\sigma(l)\sigma(m))$ with $\sigma(l)=\langle \epsilon_l^2\rangle-\langle \epsilon_l\rangle^2$ such that $Cor(l,m) \in [-1,1]$. The right bottom panel of \Fig{corr} shows that the number of pairs $(l,m)$ having $-0.3< Cor(l,m)<0.6$ increases significantly as the range of interaction increases though probability to have $ Cor(l,m)<-0.3$ or $Cor(l,m)>0.6$ decreases. When analyzed in terms of the Hamming distance between various pairs of the Fock state configurations, we see that for systems with longer range interactions there is a clear enhancement of $Cor(x)$ for the states separated by larger Hamming distances ($x > L/2$) though states separated by smaller Hamming distance also show slight increase of $Cor(x)$. Further, as shown in the top right panel of \Fig{corr},  $Cor(x) \sim a(1-x/L)^b$ with $b$ increasing from $1$ to $1.5$ as the range of interaction increases which is consistent with earlier studies on the model with NN interaction~\cite{Subroto_corr}.  Note that power-law interactions with uniform coefficients do not have any effect on correlation among local energies as expected.

So far we presented correlation for a fixed value of $h=t_0$ and $V_{ij} \in [-1,1]$. Keeping the interaction strength fixed as we increase the disorder strength, probability of having smaller $|Cor(l,m)|$  decreases though it becomes more probable to have larger values of $|Cor(l,m)|$. Eventually as $h$ keeps increasing, the distribution approaches that for the non-interacting problem $V=0$, which is peaked at $Cor(l,m) = \pm 1$, as shown in \Fig{corr2}. The exponent $b$ in $Cor(x)\sim (1-x/L)^b$ decreases and approaches $1$ as $h$ increases. For $\alpha=0.5$, this happens around $h/t_0=10$ though for shorter range interactions it happens at much smaller values of $h$. For example, for $\alpha=2$ the distribution of $Cor(l,m)$ for $h/t_0=6$ is almost same as that for the non-interacting case.
Thus, on one side random power-law interactions increase the effective disorder in the local energies but more importantly they modify correlations among local energies of the Anderson model.
\begin{figure}[h!]
  \begin{center}
    \hspace{-1cm}
\includegraphics[width=3.0in,angle=0]{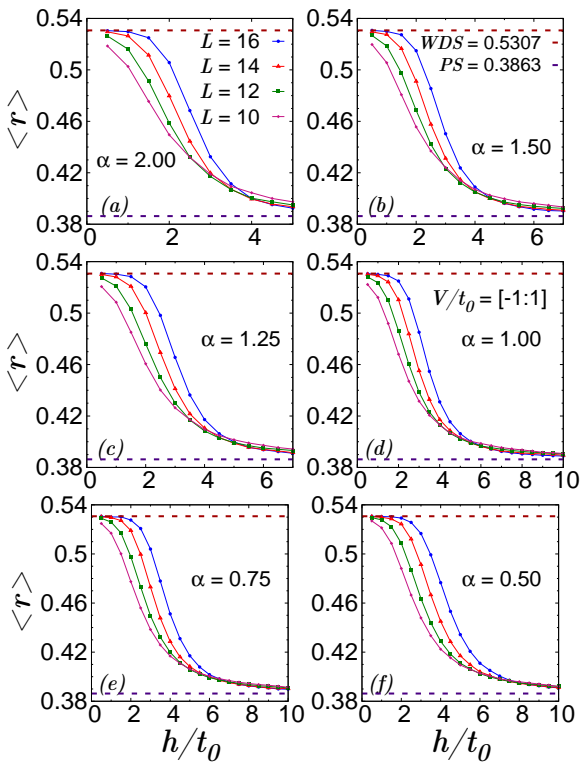}
\caption{Average level spacing ratio of successive gaps $\la r \ra$ vs disorder strength $h/t_0$ for various values of $\alpha$. The data has been averaged over $20000-500$ configurations for $L = 10-16$. We observe that below a threshold disorder, $\la r \ra$ approaches the WDS value as $L \rightarrow \infty$ indicating an ergodic state whereas above that disorder strength $\la r \ra$ approaches the PS value as $L \rightarrow \infty$ indicating a non-ergodic state. Note that curves for various $L$ do not cross at one value of $h$.}
\label{rn}
\vskip-1cm
\end{center}
\end{figure}
As we will see in further analysis, that the correlation effects among local energies dominate the physics of this model. A stronger disorder, $h$, is required to localize the system in the presence of longer range random interactions. The threshold value of $h$ at which the distribution of $Cor(l,m)$ merges with that for the non-interacting system is close to the transition point obtained from the level spacing statistics as discussed in the next section.

In the following sections, we describe various quantities that have been analysed in order to obtain the phase diagram in \Fig{pd}. The model in \Eqn{eqn:model} is solved using exact diagonalization for system sizes $L=10-18$ and various physical quantities are averaged over a large number of independent  disorder configurations $20000- 200$ respectively. Quantum quench analysis has been done using Chebyshev polynomial method for $L=16-24$ and the data presented has been averaged over (500-210) independent disorder configurations. All the results shown below are for the half-filled system for $V_{ij} \in [-1,1]$ and $t_0=1$.

\begin{figure}[h!]
  \begin{center}
\includegraphics[width=3.5in,height=4.5cm,angle=0]{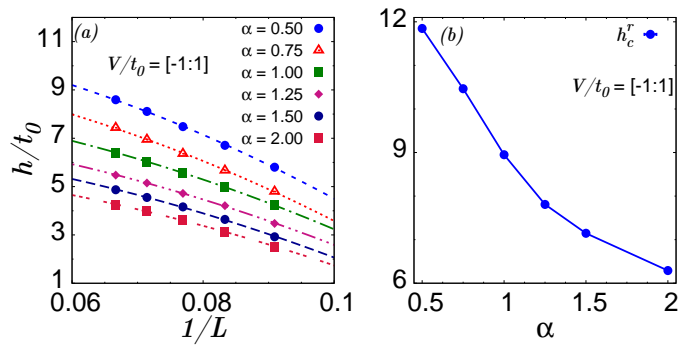}
\vskip-0.1cm
\caption{ Panel (a) shows the value of $h$ at the crossing points of $\la r \ra$ vs $h$ curves of \Fig{rn} between $L$ and $L+2$ and $L$ and $L+4$ vs inverse of the average size. The value of $h$ at the crossing points increases with increase in $\alpha$ and the system size. Dashed lines are the polynomial fits to the data. Panel $(b)$ shows the critical disorder $h_c^r$ vs $\alpha$ where $h_c^r$ is obtained by extrapolation of the fits in the thermodynamic limit.} 
\label{hc}
\vskip-1cm
  \end{center}
  \vskip-0.6cm
\end{figure}
\section{IV. Level Spacing Statistics}
We characterize the ergodic to non-ergodic transition using the eigenvalue statistics of the Hamiltonian in \Eqn{eqn:model}. The distribution of energy level spacings is expected to have Poisson statistics (PS) for a non-ergodic phase which indicates the absence of level repulsion while for an ergodic phase it is expected to follow the Wigner-Dyson statistics (WDS). We calculate the disorder averaged ratio of successive gaps in energy levels $r_n=\frac{min(\delta_n,\delta_{n+1})}{max(\delta_n,\delta_{n+1})}$ with $\delta_n=E_{n+1}-E_{n}$. The disorder averaged value of $r$ is $0.386$ for the PS; while for the WDS, the mean value of $r\approx 0.53$.  

\Fig{rn} shows the level spacing ratio $\la r\ra$ averaged over the entire many-body spectrum and over many independent realizations of disorder for various values of $\alpha$. For low disorder strength, $\la r \ra$ approaches the value expected for WDS as $L$ increases for all values of $\alpha$ studied whereas for larger disorder values $\la r \ra$ approaches the PS as $L$ increases. Interesting point to notice is that the disorder strength above which $\la r \ra$ approaches the PS value, is larger for smaller values of $\alpha$. As can be gathered from \Fig{rn}, the curves for different $L$ values do not cross at one value of $h$, but the curves for $L$ shows crossings with those for $L+2$ and $L+4$ at different values of $h$. 

\begin{figure}[h!]
  \begin{center}
    \hspace{-1cm}
\includegraphics[width=3.5in,angle=0]{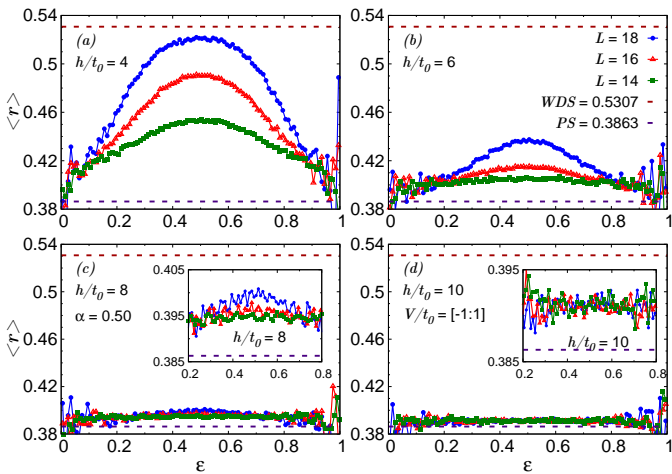}
\caption{Average level spacing ratio of successive gaps
    $r(\epsilon)$ vs normalized energy $\epsilon$ for $\alpha = 0.5$. Each panel shows data for a fixed disorder strength $h$ and three different system sizes. For $h=4t_0$ a large fraction of states in the middle of the spectrum are ergodic, with $r(\epsilon)$ approaching the value for WDS as $L$ increases though states on the edges of the spectra obey PS. As $h$ increases, the width of ergodic region of the spectrum decreases such that at $h=10t_0$, almost all the many-body states show PS for all the system sizes studied.}
\label{rvse}
\vskip-0.5cm
\end{center}
\end{figure}
To estimate the critical disorder, $h_c^r$, at which the system undergoes ergodic to non-ergodic transition, we determine the crossing points of the $\la r \ra$ curves for  $L$ and $L+2$ as well as $L$ and $L+4$. The left panel of \Fig{hc} shows these crossing points as a function of inverse of the system size for various values of $\alpha$. The data has been best fitted with the polynomial function and the extrapolated value in the limit $L \rightarrow \infty$ provides estimate for the critical disorder $h_c^r$ in the thermodynamic limit which is shown in the right panel of \Fig{hc}. Note that the value of level spacing at the crossing points approaches $0.39$ in the thermodynamic limit for any value of $\alpha$. The critical disorder $h_c^r$ increases with the range of random power-law interactions. We would like to further emphasize that though $h_c^r$ increases as $\alpha$ decreases, it stays finite even for $\alpha <1$, which is in contrast to the prediction based on resonance count argument for a system of a few particles~\cite{Dipole,Burin}.

The fact that the critical disorder $h_c^r$ increases with range of the random power-law interactions, indicates that turning on random long-range interactions delocalizes at least a part of the spectrum introducing mobility edges in the many-body eigen-spectrum. To ensure the existence of many-body mobility edges, we plot energy resolved level spacing ratio $r(\epsilon)$ as a function of normalized energy $\epsilon$ for each value of $\alpha$ and $h/t_0$. The normalized energy, $\epsilon$, is defined as $\epsilon = \frac{E-E_{min}}{E_{max}-E_{min}}$ where $E$ is the bare eigen energy and $E_{min/max}$ are the minimum and maximum eigenvalues. \Fig{rvse} shows $r(\epsilon)$ for $\alpha=0.5$ for various values of $h$ and three system sizes. In weak disorder regime, for a large fraction of states in the middle of the spectrum $r(\epsilon)$ approaches the WDS value as $L$ increases though at the edges of the spectrum $r(\epsilon)$ shows PS value. As $h$ increases, the range of $\epsilon$ for which $r(\epsilon)$ approaches the WDS reduces, such that at $h=10t_0$ the entire spectrum obeys PS. By calculating the crossing points between the curves for different system sizes we determine $E_1$ and $E_2$ such that the states below $E_1$ and above $E_2$ have PS whereas the states in between obey WDS in the thermodynamic limit. Plots of $r(\epsilon)$ for some other values of $\alpha$ are shown in Appendix A.

\begin{figure}[h!]
\begin{center}
\includegraphics[width=3.0in,angle=0]{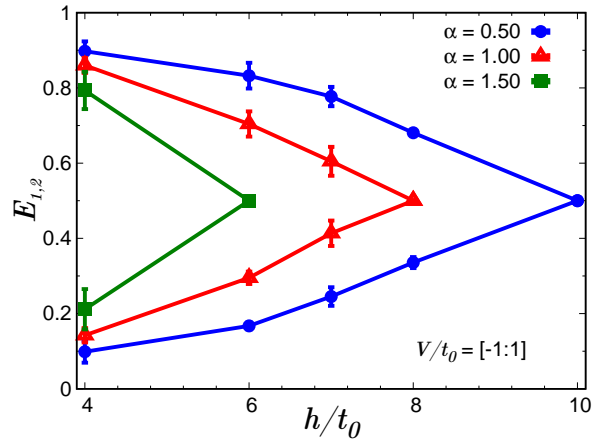}
\caption{Many-body mobility edges $E_{1,2}$ vs $h$ for various values of $\alpha$ obtained from the level spacing statistics. Many-body states below $E_1$ and above $E_2$ are non-ergodic obeying PS while the states in the middle are ergodic with WDS. For a given disorder strength $h$, the fraction of ergodic states is larger for smaller values of $\alpha$.}
\label{ME}
\end{center}
\end{figure}
\Fig{ME} shows the many-body mobility edges as a function of the disorder strength $h$ for various values of $\alpha$. At a given disorder strength, $h$, the range of energy over which many-body states remain ergodic increases as $\alpha$ decreases.
The fraction of non-ergodic states increases with the increase in the disorder strength $h$ and eventually the mobility edges disappear after a critical disorder strength $h_{ME}$, which is a function of $\alpha$. The critical value $h_{ME}$ is larger for smaller values of $\alpha$. Note that $h_{ME}$, obtained from analysis of energy resolved level spacing for $L$ up to $18$ is consistent with $h_c^r$, the critical disorder in the thermodynamic limit.

\section{V. Time Evolution of Density Imbalance}
We study dynamics of the system after a quench starting from an initial charge density wave (CDW) ordered state $|\Psi_0\ra =\prod_{i=0}^{L/2-1} C^\dagger_{2i}|0\ra$ and calculate the time evolution of the density imbalance between even and odd sites to distinguish between the localized and delocalized phases. Imbalance $I(t)$, for the initial state considered, is defined as 
\be
I(t) = \frac{\sum_{i=0}^{L-1} (-1)^i\la n_i(t)\ra}{\sum_{i=0}^{L-1} \la n_i(t) \ra}
\ee
\begin{figure}[h!]
  \begin{center}
\includegraphics[width=3.4in,angle=0]{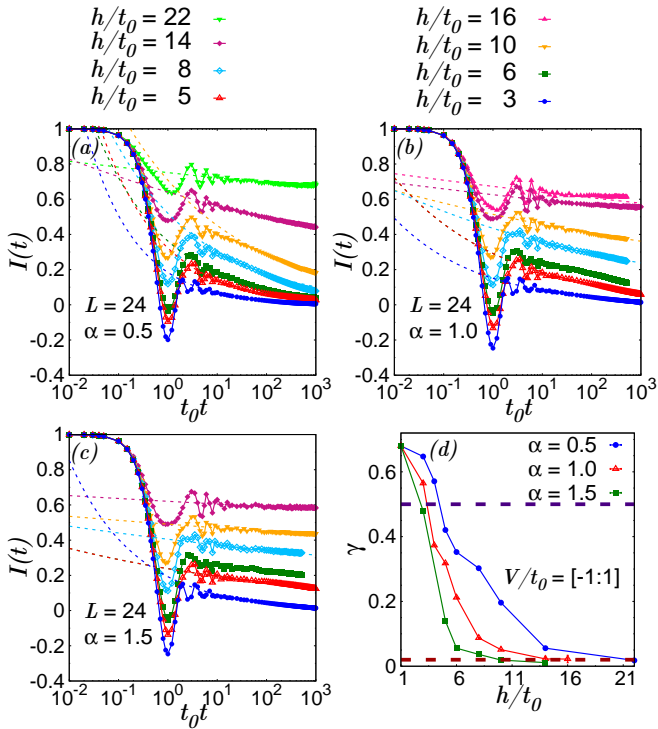}
\caption{The density imbalance $I(t)$ as a function of time for various values of disorder $h$ and $\alpha$ for $L=24$. The points are the results of numerical calculation using Chebyshev polynomial method and lines are the fit to the form $t^{-\gamma}$. For very small disorder values, for example, $h=3t_0$, $I(t)$ decays with the exponent $\gamma >1/2$ for $\alpha <1$ representing a super-diffusive regime though for $\alpha=1.5$ the system is almost diffusive with $\gamma \sim 1/2$. As $h$ increases, the rate of decay of the imbalance decreases resulting in $\gamma <1/2$. The value of disorder $h_c^I$ at which the imbalance shows saturation and no significant power-law decay (basically $\gamma \le 0.01$) is much larger for smaller values of $\alpha$ resulting in a broad sub-diffusive phase above $h_c^r$ for longer-range interacting systems. Panel (d) shows the plot of $\gamma$ vs $h$ for three $\alpha$ values studied. The dotted lines are cuts at $\gamma=0.5$ corresponding to the diffusive phase and $\gamma=0.01$ corresponding to the MBL phase.}
\label{imbalance}
\end{center}
\end{figure}
Density imbalance is a signature of how much memory the system has of the initial order after certain time steps and can be easily probed in experiments~\cite{expt2}. Starting from an initial state $|\Psi_0\ra$, we let the state evolve w.r.t the Hamiltonian in \Eqn{eqn:model} to obtain the time evolved state $|\Psi(t)\ra=exp(-iHt)|\psi_0 \ra$ and calculate $I(t)$ as a function of time which is then averaged over many independent disorder realizations. Time evolution is carried out numerically using Chebyshev polynomial method~\cite{Weiss,Fehske,Holzner,Halimeh,Soumya}, details of which are given in Appendix B. The density imbalance has an initial rapid decay followed up by oscillations. For intermediate to large time regime, $I(t)$ shows a power-law decay superimposed on decaying oscillations $I(t) \sim t^{-\gamma}$~\cite{Yevgeny_rev}. The exponent $\gamma$ has been shown to be related to the dynamical exponent of the mean square displacement $\la x^2 \ra \sim t^{2/z}$ as $\gamma =1/z$~\cite{exponent}.

\begin{figure*}
  \begin{center}
   \hspace{-1cm}
\includegraphics[width=5in,angle=0]{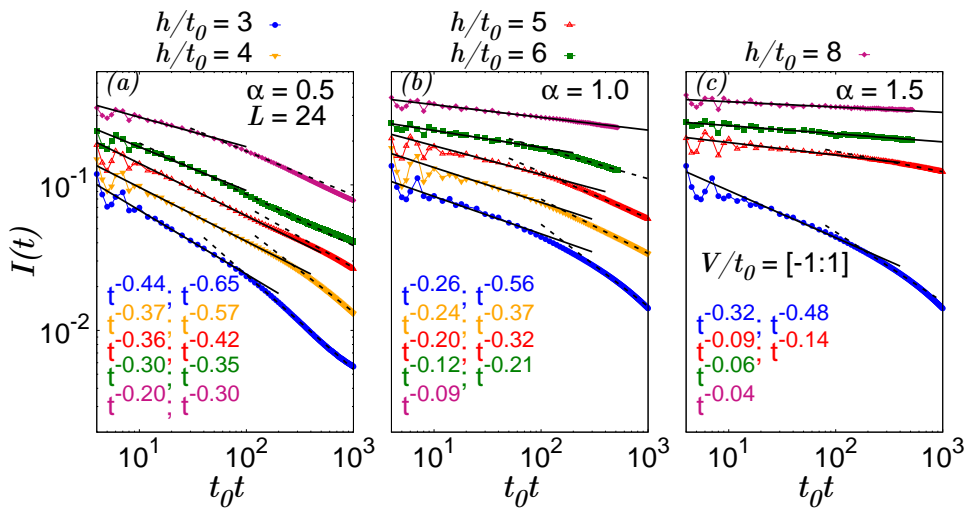}
\caption{Zoom-in plots of the density imbalance $I(t)$ as a function of time on log scale for various values of disorder $h<h_c^r$ and $\alpha$. For smaller values of $h \ll h_c^r$, $I(t)$ clearly shows a change of slope after $t \sim 100$ such that $\gamma$ obtained by fitting $I(t)$ for shorter time range (mentioned first in the keys) is smaller than the $\gamma$ value obtained from the fit for $t>100$ (written later in the keys). As $h$ increases, the two $\gamma$ values become closer to each other and for $h>h_c^r$ we could not see a significant slope change in the imbalance. The data shown is for $L=24$.}
\label{I_fits}
\end{center}
\end{figure*}

\Fig{imbalance} shows $I(t)$ for various values of disorder and $\alpha <2$. For a fixed disorder strength, $h$, the imbalance shows the fastest decay for the smallest value of $\alpha$. This implies that the system with longer range interaction has less memory of the initial state and hence is more ergodic which is consistent with the analysis of level spacing statistics and the correlation among local energies in the Fock space. A more quantitative analysis of the imbalance can be done by fitting the imbalance data to the power-law decay form $I(t)\sim t^{-\gamma}$ for intermediate to large time regime. The dashed lines in \Fig{imbalance} show the fits to the data and zoom-in fits are shown on log-scale in \Fig{I_fits}. Firstly, as shown in \Fig{I_fits}, for smaller values of $h$, the imbalance shows a change of slope around $t_0t \sim O(100)$ such that the exponent $\gamma$ derived from the long time fit is larger than the $\gamma$ obtained from fits upto $t_0t\sim 100$. As $h$ increases the difference between the two values of $\gamma$ reduces and eventually for $h \ge h_c^r(\alpha)$ the imbalance data in the entire time range studied can be fitted with one exponent. We believe that the change in dynamical exponent with time for $h < h_c^r$  is an indication of more than one relaxation time scales in the system due to the presence of both ergodic and non-ergodic states, which are separated by the many-body mobility edges.
In panel(d) of \Fig{imbalance}, we have shown the larger $\gamma$ values for $h < h_c^r$ where change of slope occurs.

There are a couple of important observations to be made from \Fig{imbalance}. Firstly as shown in panel (d), for very small values of disorder $h\le 2t_0$, the density imbalance shows super-diffusive dynamics with $\gamma >1/2$ for all values of $\alpha$ studied. The corresponding imbalance plots are shown in Appendix B. For longer range interactions ($\alpha < 2$) the super-diffusive regime gets extended up to $h_0$ beyond $2t_0$ as shown in \Fig{I_fits}. For $\alpha=0.5$ the imbalance shows super-diffusive behaviour up to $h=4t_0$ while for $\alpha=1.0$, $\gamma$ remains more than $1/2$ up to $h=3t_0$. In systems with quasi-periodic potential, super-diffusive transport has been observed earlier in the non-interacting case~\cite{Dhar} as well as for MBL systems with weak disorder and weak interactions~\cite{Yevgeny-AA}. This is because the non-interacting system with quasi-periodic potential has extended states that are ballistic and not diffusive~\cite{Huse2013,AA} and even in the presence of weak interactions the extended states remain super-diffusive. 
But in the model we studied, though for $h <2t_0$ the system has single particle mobility edges separating the extended (ballistic) states from the localized states, the half-filled interacting system is fully ergodic and extended~\cite{garg} resulting in superdiffusive transport. For longer range interactions, even for $h = 4t_0$ more than $99\%$ of the many-body states are ergodic and extended and hence the system continues to have superdiffusive transport. Compared to earlier studies on quasi-periodic potential~\cite{Yevgeny-AA}, the super-diffusive phase in our model appears for much larger strength of interactions $V_{ij}\in [-1,1]$ and the aperiodic potential $h$ and gets broadened for longer range of random power-law interactions. This is because of enhanced delocalization of many-body states due to power-law interactions with random coefficients and is in complete consistency with our level spacing analysis.
We believe that the system must have diffusive dynamics with $\gamma \sim 1/2$ at least for a narrow window above $h_0$, though in our numerics we did not see $\gamma$ exactly being equal to $1/2$ for any of the parameter values studied.

As $h$ increases further but still staying below $h_c^r$, the system which has many-body mobility edges separating the localized non-ergodic states from the ergodic states in the middle of the spectrum, enters into the slow dynamics sub-diffusive phase with $\gamma < 1/2$. The exponent $\gamma$ decreases monotonically with increase in $h$, as shown in \Fig{imbalance} and \Fig{I_fits}, which can be explained in terms of slowly decreasing fraction of the ergodic many-body states with increase in the disorder strength.  Ergodic sub-diffusive phase has been observed in many earlier works, both, theoretically~\cite{Yevgeny_rev,Yevgeny-AA,Soumya,Luitz,Znidaric} and experimentally~\cite{dynamics_expt} for systems with nearest-neighbour interactions in the presence of random as well as quasi-periodic potential. We observe a broad ergodic sub-diffusive phase for $h_0 < h < h_c^r$ even in the presence of long range interactions in this model and the width of this phase increases for longer-range interactions. We discuss the possibility of Griffiths effects behind the sub-diffusive phase in details in section VII.

Furthermore, even for $h>h_c^r$, where the system is fully non-ergodic with the level spacing ratio showing PS for the entire many-body spectrum, the dynamics continues to be sub-diffusive with $0 < \gamma <1/2$ for a wide range of disorder strength. As the disorder increases further, the system transits into the MBL phase with $\gamma \le 0.01$ at $h_c^I(\alpha)$ as shown in Fig.~\ref{imbalance} and also in the phase-diagram of \Fig{pd}.
The width of the non-ergodic sub-diffusive phase is significantly large for $\alpha <1$ resulting in larger value of $h_c^I$ at which the system enters into the MBL phase. The broad non-ergodic sub-diffusive phase observed in this analysis, preceding the MBL phase, is analogous to the delocalized non-ergodic phase or  ``bad metal'' phase  proposed to exist in short range interacting systems for disorder strengths below the MBL transition~\cite{Altshuler,Serbyn}. This will become more clear from the analysis of eigenfunction statistics in the next section. However, there is no consensus so far about the fate of this ``bad metal'' phase in the thermodynamic limit. Whether this phase shrinks in the thermodynamic limit to a critical point or remains of finite width in the parameter space is an open question~\cite{Santos,Serbyn}. According to a recent theoretical work, if the non-interacting system has single particle mobility edges, in the corresponding interacting system the non-ergodic sub-diffusive phase may persist even in the thermodynamic limit~\cite{Subroto-NEE}. In our model, we observe a broad non-ergodic sub-diffusive phase for $h \gg 2t_0$, where all the single-particle states are highly localized. Also the range of this phase gets broader as the range of random power-law interactions increases.

\begin{figure}[h!]
\begin{center}
\includegraphics[width=3.3in,angle=0]{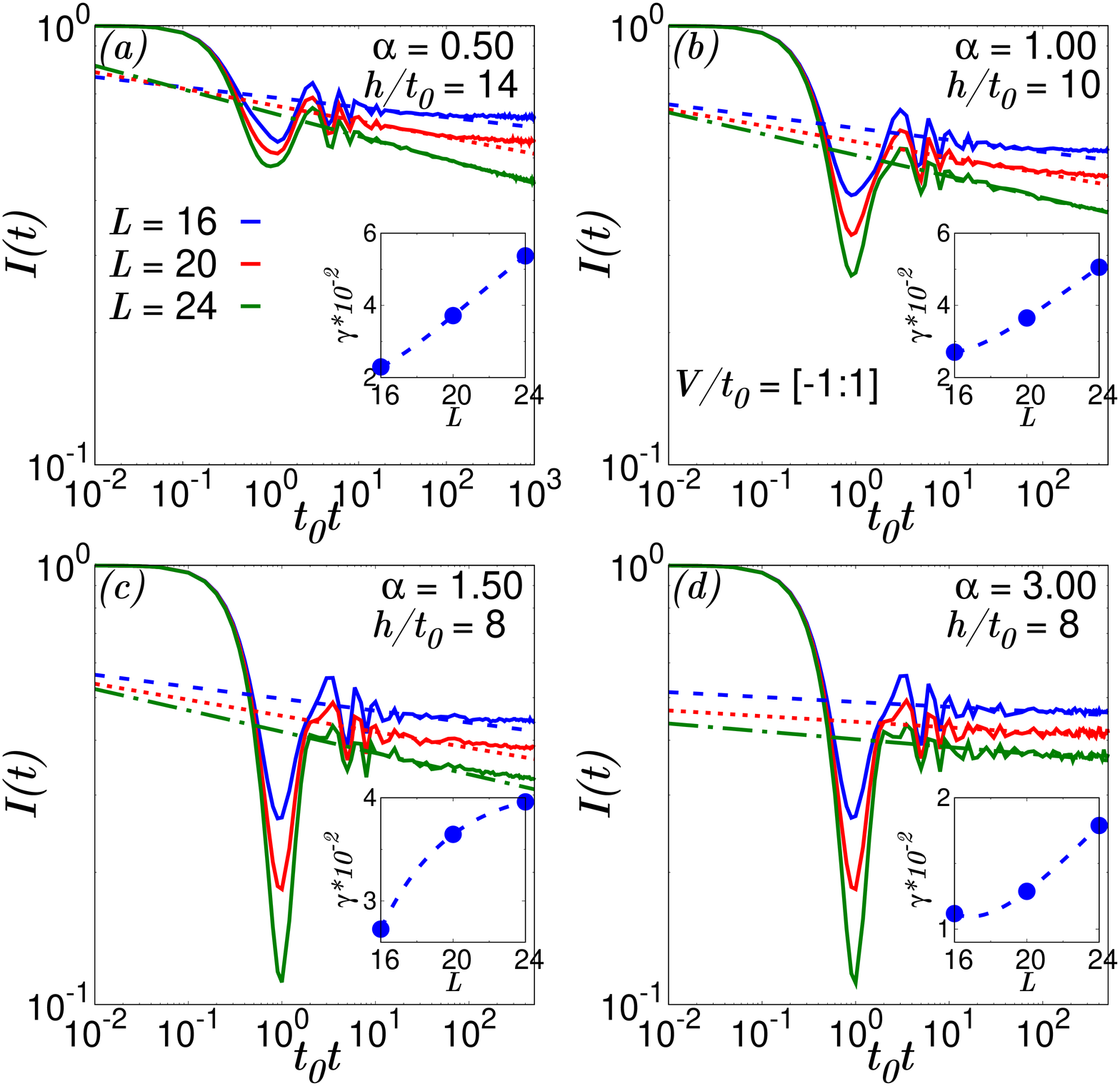}
\caption{The density imbalance $I(t)$ as a function of time $t$ for three different system sizes for various values of $\alpha$. The $h$ values are chosen such that it is slightly larger then or close to $h_c^r$. Though for $\alpha=3$, there is hardly any increase in $\gamma$ with the system size, a clear increase in $\gamma$ with $L$ is observed for smaller $\alpha$ values.}
\label{I_L}
\vskip-0.2cm
\end{center}
\vskip-0.7cm
\end{figure}

Now we focus on the system size dependence of the imbalance. The imbalance data shown so far is for $L=24$ sites chain. But there is a significant system size dependence of the density imbalance as shown in \Fig{I_L}. As the system size increases, the imbalance shows a faster decay with a larger exponent $\gamma$. This has been observed earlier in systems with short-range interactions~\cite{Soumya,Mirlin-imbalance} though the system size effect was found to be less serious for models with quasi-periodic potential compared to those with fully random potential~\cite{Mirlin-AA}. Although the model we have studied has aperiodic potential, which is very close to the quasiperiodic potential studied in~\cite{Mirlin-AA}, but we also have random power-law interactions. In fact, the increase in $\gamma$ is more significant for the system with longer range of interaction. Given this, we can not rule out if the entire ergodic sub-diffusive phase or at least a significant part of it actually turns out to be diffusive in the thermodynamic limit.  However, given the small values of $\gamma$ for $h_c^r < h < h_c^I$, the non-ergodic sub-diffusive regime will probably still remain robust with minor modifications as the system size increases.
Secondly, our estimate of the disorder strength $h_c^I$ above which $\gamma <0.01$ and the system enters into the MBL phase, will also  shift upwards due to increase in $\gamma$ in the thermodynamic limit. This would further broaden the non-ergodic sub-diffusive phase.

Generally, the sub-diffusive phase near the MBL transition is associated with multifractality of the eigenstates~\cite{Yevgeny_rev,Yevgeny-fractal}. In the next section we analyse eigen-function statistics in order to develop understanding of the mechanism of the sub-diffusive phases observed in this system.

\section{VI. Eigen-function Statistics}

Eigenfunction statistics has played a crucial role in understanding of Anderson localization~\cite{Mirlin_rev} as well as many-body localization~\cite{Alet_rev}. In the non-interacting disordered system, eigenfunctions have been shown to be multifractal near the Anderson transition~\cite{Mirlin_rev}, which means that the eigenfunctions are neither extended nor localized but cover a sub-extensive number of sites. Similarly, close to the single particle mobility edges, eigenfunctions have been shown to have multifractal behaviour~\cite{fractal-ME}. Below, we calculate inverse participation ratio (IPR) and higher moments in order to analyse many-body eigenfunctions.

Given an eigenstate $|\Psi_n\ra = \sum_l \Psi_n(l)|l\ra$, in the basis state $|l\ra$, $q^{th}$ moment is defined as  $I_q(n)=\sum_{l=1}^N |\Psi_n(l)|^{2q}$ such that for $q=2$, $I_q$ gives the IPR which measures the extent of delocalization of the eigenstate $|\Psi_n\ra$ in the basis $|l\ra$. Here we chose $|l\ra$ to be the basis in the Fock space of spin-less fermions and $N$ is the dimension of the Fock space. An extended state, which gets contribution from almost all the basis states of the Fock space has $IPR(n)\propto 1/N$ while for a localized state $IPR(n) \propto O(1)$ in the thermodynamic limit. There is a third intermediate phase possible, which is known as multifractal phase~\cite{Santos,Serbyn} for which IPR goes to zero in the thermodynamic limit as $IPR(n)\propto 1/N^\mu$ but with $\mu <1$ and in general the $q^{th}$ moment goes as $I_q(n)\propto 1/N^{\mu_q(q-1)}$ with the generalized fractal dimension $\mu_q$ deviating from one and having nonlinear dependence on $q$. For discussion below, we have dropped the subscript for IPR ($q=2$). This phase is also known as non-ergodic (having biased contribution from basis states in the Fock space) extended phase. It is important to note that fractal properties strongly depend on the choice of basis. In this work, we chose the standard basis of product states in the Fock space,  like in many of the previous works~\cite{Santos,Yevgeny-fractal}.

\begin{figure*}
\begin{center}
\includegraphics[width=4.5in,angle=0]{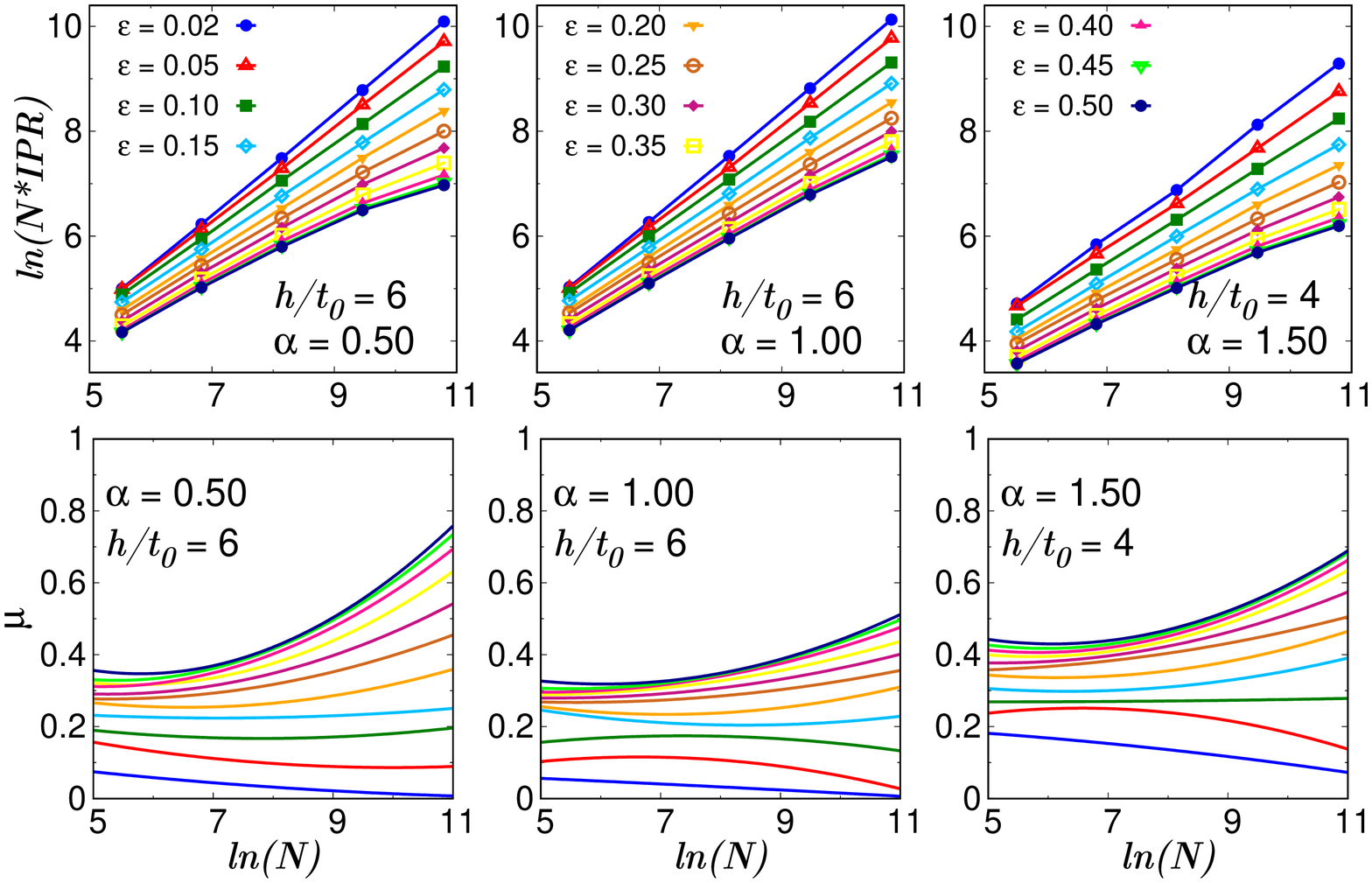}
\caption{Inverse participation ratio $IPR$ for various values of the normalized eigen-energy $\epsilon$ and $\alpha$. $h$ values are chosen such that $h < h_c^r$. The top panels show $\ln(N*IPR(N))$ vs $\ln(N)$ where $N$ is the dimension of the Fock space. The bottom panels show the fractal exponent $\mu$ as a function of $\ln(N)$ for various values of $\epsilon$ and three values of $\alpha$. Note that states on the edges $\epsilon \le 0.05$ are localized with $\mu \rightarrow 0$ in the thermodynamic limit while states in the middle are extended with $\mu$ increasing as $N$ increases. The intermediate states $0.05 < \epsilon < 0.2$ (for $\alpha=0.5,1.0$) are mutifractal with $\mu \ll 1$.} 
\label{IPR-e}
\end{center}
\end{figure*}

First we analyse IPR for the entire many-body spectrum for $h_0 < h < h_c^r$ where the system shows many-body mobility edges from the level spacing statistics. \Fig{IPR-e} shows $IPR(\epsilon)$ for various values of the normalized energy $\epsilon$ for three $\alpha$ values. Following~\cite{Mirlin-fractal}, we have analysed the scaling of $\ln(N*IPR)$ vs $\ln(N)$ and the flowing fractal exponent $\mu = -\frac{\partial \ln(IPR(N))}{\partial \ln(N)}$, both shown in \Fig{IPR-e}. Panel (a,b) shows the data for $h=6t_0$ for $\alpha=0.5$ and $1.0$. For states near the edges of the spectrum, for example $\epsilon \le 0.05$ for $\alpha=0.5$ and $\epsilon \le 0.1$ for $\alpha=1.0$, $N*IPR$ increases linearly with $N$ for large $N$ which is a signature of the localised state. As an effect $\mu \rightarrow 0$ in the thermodynamic limit. For states in the middle of spectrum, with $0.2 \le \epsilon \le 0.5$, $N*IPR$ saturates for $N \gg 1$ and hence $\mu$ approaches $1$ in the thermodynamic limit. But for the states close to the mobility edges of Fig.~\ref{ME}, that is in the range $0.05 < \epsilon <0.2$, the fractal exponent $\mu$ neither increases in the thermodynamic limit nor it is vanishingly small, but it remains non-zero and much less than 1, indicating multifractal nature of states near the mobility edges. Similar trend is observed for $\alpha=1.5$ and  $h=4t_0$, as shown in panel (c) of \Fig{IPR-e}.

\begin{figure*}
\begin{center}
\includegraphics[width=4.5in,angle=0]{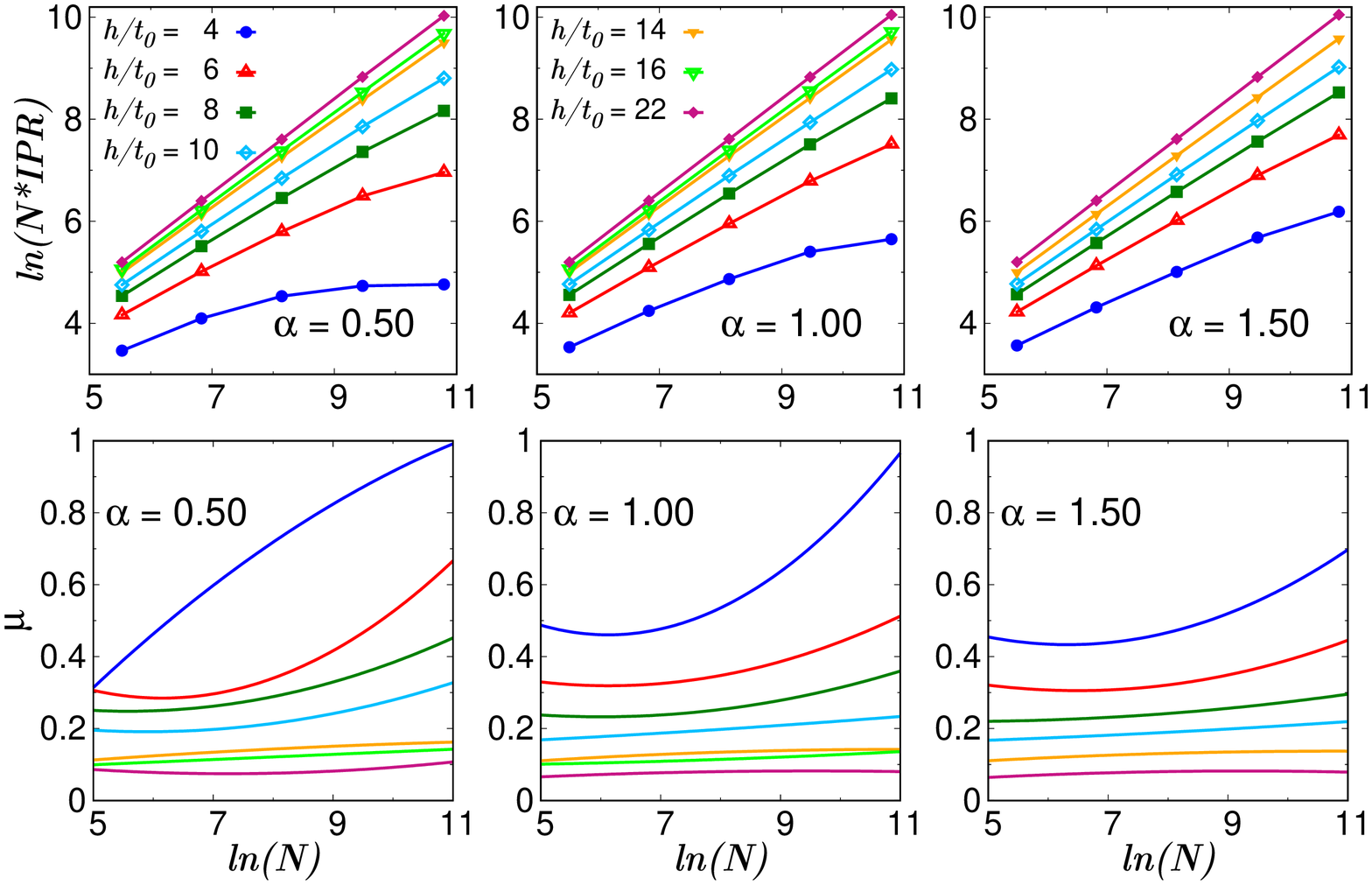}
\caption{Inverse participation ratio $IPR$ for states in the middle of the spectrum for various values of $\alpha$ and $h$. The top panels show $\ln(N*IPR(N))$ vs $\ln(N)$ where $N$ is the dimension of the Fock space. The bottom panels show the fractal exponent $\mu$ as a function of $\ln(N)$ for various values of $h$ and three values of $\alpha$.}
\label{IPR}
\end{center}
\end{figure*}

Next we study eigenfunctions in the middle of the spectrum for various values of $h$, as shown in \Fig{IPR}. We calculate $IPR$ for $1/10$th of the states in the middle of the spectra, that is, around $\epsilon=0.5$ for various system sizes $L=10-18$ and average it over many independent disorder realizations. For very large values of $h$, for example $h=22t_0$, $N*IPR$ increases linearly with $N$ and $\mu \sim 0$ for large $N$, which is a signature of the localized state though $\mu$ increases a bit as $\alpha$ decreases. On the other hand, for conventional extended states $N*IPR$ should saturate for $N \gg 1$ to a value which increases with the disorder strength. For $\alpha=0.5(1.0)$, we see this trend for $h \le 6t_0$ ($h\le 4t_0$) though for larger $\alpha$ values this trend is seen for further smaller values of $h$. In this regime, the fractal exponent $\mu$ approaches its conventional value $1$ as the Fock space volume increases, as seen clearly in the panels for $\alpha=0.5$ and $1.0$ for $h=4t_0$.
For intermediate values of disorder, like $6<h/t_0<10$ for $\alpha=0.5$ and $4<h/t_0<8$ for $\alpha=1.0$, though $N*IPR$ does not show saturation for the system sizes studied, but the fractal exponent $\mu$ shows an increasing trend in the large $N$ limit indicating that the states are extended here as well. Thus for $h <h_c^r$, where the mid-spectra level spacing ratio obeys WDS, the corresponding eigenstates are also extended in the Fock space. But for $h_c^r < h < h_c^I$, the fractal exponent $\mu$ neither increases in the thermodynamic limit nor it is vanishingly small but $\mu$ remains non zero, being much less than one indicating the multifractal nature of the eigenstates.
\begin{figure}
  \begin{center}
    \hspace{-1cm}
\includegraphics[width=3.5in,angle=0]{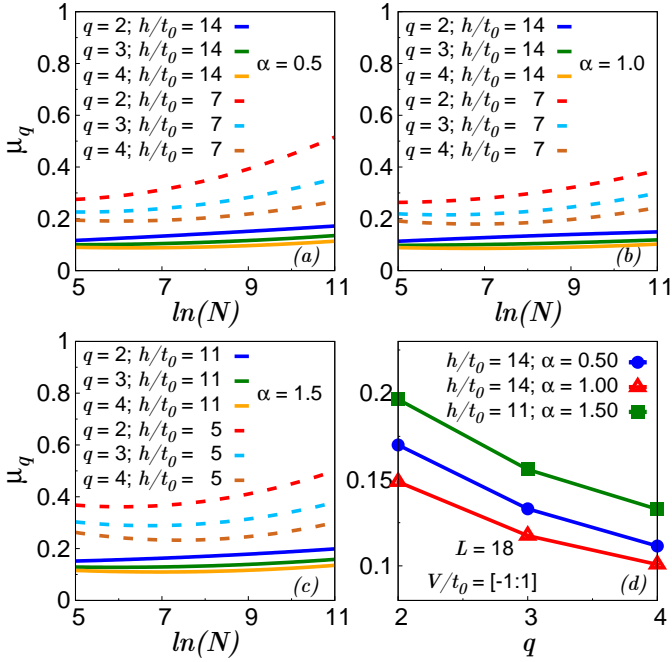}
\caption{Generalized fractal dimension $\mu_q$ for $q=2,3,4$ for states in the middle of the spectrum for various values of $\alpha$ and $h$. For $h<h_c^r$ $\mu_q$ increases with the system size for all $q$ studied and for all values of $\alpha$, while for $h_c^r < h < h_c^I$ $\mu_q \ll 1$ for all values of $q$ and $\alpha$ studied here. Panel (d) shows the dependence of generalized fractal exponent $\mu_q$ on $q$ for various values of $h$ and three values of $\alpha$.}
\label{muq}
\end{center}
\end{figure}

At the end, we discuss the higher moments for $h <h_c^r$ as well as $h>h_c^r$ for $q=3$ and $4$. \Fig{muq} shows the generalized fractal dimension $\mu_q$ for the above mentioned values of $q$ for the states in the middle of the spectrum. For $h< h_c^r$, $\mu_q$ increases with the system size and should approach unity in the thermodynamic limit, supporting our conclusion from IPR analysis that the states are fully extended in this parameter regime. But for $h_c^r < h< h_c^I$, $\mu_q$ does not show clear increase with the system size and $\mu_q \ll 1$ for all the values of $q$ studied. Panel (d) of \Fig{muq} shows a nonlinear dependence of $\mu_q$ on $q$ for $h_c^r < h < h_c^I$ which confirms multifractal nature of eigenstates.

To summarise, quantum quench dynamics showed two regimes of sub-diffusive transport, one below $h_c^r$, where the system has many-body mobility edges and the other for $h_c^r < h < h_c^I$. Eigenfunction statistics reveals that for $h < h_c^r$ the many-body states near the mobility edges are multifractal though the states in the middle of the spectrum are extended. For $h_c^r < h \le h_c^I$, the eigenstates in the middle of the spectrum are multifractal.

\section{VII. Conclusions and Discussions}
In this work, we have explored competing effects of the long-range power-law interactions $V_{ij}r_{ij}^{-\alpha}$ with random coefficients $V_{ij}$ and the one body aperiodic potential, $h$. We demonstrated that $h$ required to attain the MBL phase in the presence of random power-law interactions is much larger compared to the case of short range interactions, which we attribute to the modification in correlation among local energies of the effective Anderson model on Fock space due to random long-range interactions. The MBL phase in the presence of random long-range interactions is preceded by a broad non-ergodic sub-diffusive phase whose width increases with decrease in $\alpha$. The non-ergodic sub-diffusive phase observed here is characterized by Poissonian statistics for level spacing ratio, slow dynamics in quantum quench and multifractality of the eigenstates. Generally, the non-ergodic extended phase is expected to appear only near the MBL transition~\cite{Altshuler,Santos,Serbyn,Yevgeny-fractal} though its fate in the thermodynamic limit has been debatable. In this work we have observed a broad non-ergodic sub-diffusive phase which interestingly has parallels with the slow non-thermalizing dynamics that has been recently observed in models of quantum spin-glass~\cite{Abanin_glass} and with the ``quasi-MBL'' phase observed in an infinite range random interactions model of fermions~\cite{Thomson}.

Based on the time dynamics of a charge density wave ordered initial state, we also identify a super-diffusive phase for $h<h_0<h_c^r$, characterized by the fast decay of the density imbalance with the dynamical exponent $\gamma >1/2$. 
For $h_0 < h < h_c^r$, where the level spacing ratio shows mobility edges separating ergodic states in the middle from the non-ergodic states at the edges of the spectrum, the system is sub-diffusive with $\gamma < 1/2$ as shown in the phase diagram of \Fig{pd}. The width of this ergodic sub-diffusive phase increases for smaller values of $\alpha$, where the interactions are long-range in nature. Even for $h >h_c^r$, the dynamics remain sub-diffusive for a wide range of the disorder strength and becomes slower as the disorder strength increases. Eventually the system transits into the MBL phase which is non-ergodic and diffusion-less.

In short, in our model we  observe two regimes of sub-diffusive transport. The sub-diffusive phase closer to the MBL transition (for $h_c^r < h < h_c^I$) is completely non-ergodic and has multifractal eigenstates in the middle of the spectrum. The other sub-diffusive phase appears at much lower values of disorder, $h < h_c^r$, where the states near the many-body mobility edges are multifractal. This seems consistent with earlier proposals in which sub-diffusive transport close to the MBL phase is explained in terms of  multifractality of the eigenstates~\cite{Serbyn,Yevgeny_rev,Yevgeny-fractal,Mirlin_rev}. The question of interest is, why random long-range interactions result in multifractal behaviour of eigenstates over such a broad parameter regime and will be explored in future works.

There are suggestive explanations of the sub-diffusive transport in terms of the Griffiths phase for systems with short-range interactions in 1-d, where the rare insulating regions act as bottlenecks for transport resulting in slow dynamics of the system~\cite{Griffiths} close to the MBL transition. Though the rare region effects are absent for systems with deterministic potential, the model we have studied also has random power-law interactions which can have``rare regions''. Generally, it is believed that Griffiths effects in 1-d systems with long-range interactions are similar to those in higher dimensions with short range interactions where insulating inclusions in an ergodic phase can be bypassed. But, we would like to emphasize, that this is true only if the hopping is long range in nature. With nearest neighbour hopping and long-range interactions with random coefficients, the one-dimensional system must have rare region effects. This is better supported by comparing the transport in this model in the presence of long-range interactions with uniform coefficients rather than random coefficients. For the system with uniform long range interactions, in the non-ergodic regime, the imbalance does not show any decay with time and the system remains localized (as shown in Appendix B) though in the presence of random long-range interactions we observed a broad sub-diffusive phase. We believe that this difference in dynamics of the two systems is because of the rare region effects which are present in the system with random long-range interactions and are absent for the case of uniform power-law interactions. 
There can be rare regions where interactions are very small and hence the disorder is effectively strong, which can result in ``insulating'' bubbles in the ergodic phase for $h < h_c^r$ and may lead to sub-diffusive transport. Similarly, for $h> h_c^r$, there might be rare regions where the interactions are weak or purely repulsive which might generate extended bubbles like $|010100....\ra+|100010\ra+|100100\ra+...$ in an otherwise MBL phase. These thermal bubbles in the otherwise MBL phase prohibit complete localization of the system and result in a sub-diffusive phase near the MBL transition.

  It will certainly be interesting to explore other dynamical quantities like mean square displacement, conductivity and  return probability especially for bigger system sizes and if possible using analytic calculations. Understanding the mechanism behind wide regions of anomalous transport, and Griffiths effect in the presence of random-long range interactions is crucial and will be explored in future works. Finally, since the range of power-law interactions can now be controlled in state-of-the-art experiments~\cite{tc1,tc,ions2}, it will be interesting to explore MBL in the presence of power-law interactions with random coefficients in experiments and to look for the proposed non-ergodic sub-diffusive phase in these systems.

\section{Acknowledgements}
A.G. would like to acknowledge insightful discussions with A.D. Mirlin about MBL systems with long range interactions during the conference ``Thermalization, Many-body localization and Hydrodynamics'' (Code:ICTS/hydrodynamics2019/11) held at ICTS, Bengaluru. A.G. is thankful to A. L. Burin for email communications  about issues related to the long range interactions with random sign. A.G. also acknowledges Science and Engineering Research Board (SERB) of Department of Science and Technology (DST), India under grant No. CRG/2018/003269 for financial support. Y.P. would like to acknowledge DST for funding, and SINP cluster facilities.

\section{Appendix A}
\begin{figure}[h!]
  \begin{center}
    \hspace{-1cm}
\includegraphics[width=3.6in,angle=0]{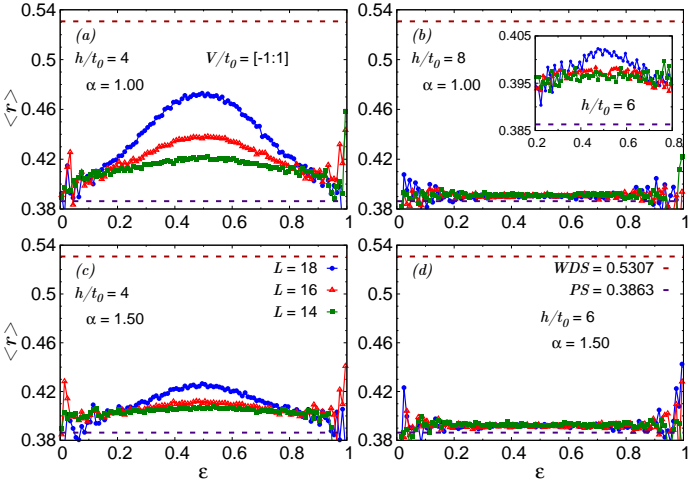}
\caption{Ratio of successive gaps $r(\epsilon)$ vs $\epsilon$ for various values of $h$ and three system sizes. For $\alpha=1.0$, the system has many-body mobility edges for $h \le 6t_0$ though the entire spectrum obeys PS for $h=8t_0$. But for $\alpha=1.5$, transition to the non-ergodic phase occurs at $h\le 6t_0$.}
\label{rint}
\end{center}
\end{figure}

In this appendix, we present energy resolved level spacing statistics for $\alpha=1.0$ and $1.5$. 
\Fig{rint} shows the plot of $r(\epsilon)$ vs $\epsilon$ for the system with random power-law interactions for various values of $h$ and three system sizes. For $\alpha=1.0$ the system shows mobility edges, which separate states obeying PS from those which obey WDS, upto $h=6t_0$. For $h=8t_0$ the system is fully non-ergodic with level spacing ratio for all the many-body states showing PS value. For shorter range interactions, the transition to non-ergodic phase occurs at smaller values of $h$.
As shown in the bottom panel of \Fig{rint}, for $\alpha=1.5$, already at $h=6t_0$, $r\sim 0.386$ for the entire many body spectrum.

\section{Appendix B}
In this section we provide brief description of the Chebyshev polynomial method used to evaluate the time evolution of the density imbalance and also some results related to the density imbalance.
\\
  {\bf{Chebyshev Polynomial Method for time evolution:}}\\
  The direct calculation for time-evolution is not feasible
for larger Hilbert-space dimension using exact diagonalization
technique which restricts the calculation to $L = 16$ system
sizes. Hence, to study dynamics in large system sizes, we use the 
Chebyshev polynomial method which is a well known and established method for time evolution of quantum systems~\cite{Weiss,Fehske,Holzner,Halimeh} and has been used to study quantum quench dynamics of MBL systems~\cite{Soumya}.

In chebyshev scheme, we express the time-evolution operator
$U(t,0)$ in terms of a finite series of first-kind Chebyshev
polynomials of order $k$.
One important point to note is that
the whole set of Chebyshev polynomials is defined on the
interval $[-1:1]$. Hence before expanding the time evolution
operator $U(t,0)$ in terms of Chebyshev polynomials, we must
shift and rescale the Hamiltonian $\tilde{H} = \frac{H-b}{a}$
to restrict the spectrum within the interval $[-1,1]$
\cite{Weiss}. The parameters
$b = \frac{1}{2}(E_{max} + E_{min})$ and
$a = \frac{1}{2}(E_{max} - E_{min} + \epsilon)$ where $E_{max}$
and $E_{min}$ represents the extreme eigenvalues of the
Hamiltonian $H$. A small parameter $\epsilon$ has been
introduced to ensure rescaled eigenvalues
$|\tilde{E}| \leq \frac{1}{1+\delta}$ lies well inside
$[-1:1]$ \cite{Fehske}. We use Lanczos method to obtain the largest and smallest eigenvalues of $H$ for $L=24$ for a few realizations of disorder configurations for various parameters in the Hamiltonian. Then we use a slight overestimation of $E_{max}-E_{min}$ for normlization of the spectrum. For practical purpose we chose
$\delta=0.01$.

\begin{figure}
\begin{center}
  \includegraphics[width=3.1in,angle=0]{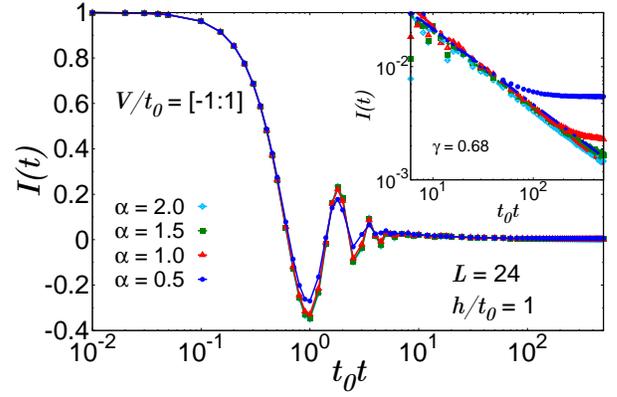}
  \caption{Density imbalance $I(t)$ vs $t_0t$ for $h=t_0$ for various values of $\alpha$. The data shown is for $L=24$.}
\label{I1}
\end{center}
 \end{figure}

\begin{figure}
\begin{center}
\includegraphics[width=2.5in,angle=0]{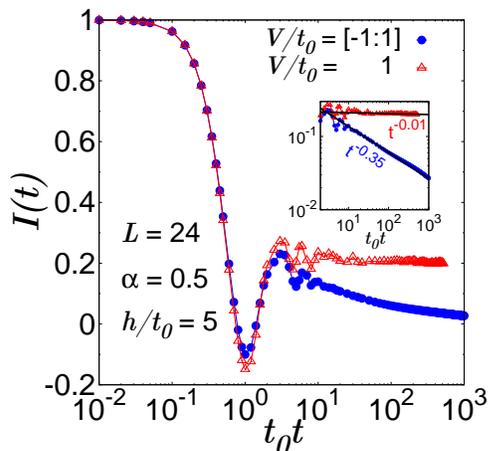}
\caption{The density imbalance $I(t)$ as a function of time for $h=5t_0$ and $\alpha=0.5$. The red curve represent the results for the system with uniform power law interactions with $V=t_0$ while the blue curve is the result for the system with random power-law interactions.}
\label{I2}
\end{center}
\end{figure}

  {\bf{Density Imbalance for $h=t_0$:}}
  \\
 For very small values of disorder $h\le 2t_0$, where the non-interacting system has single particle mobility edges, and hence the half-filled interacting system is fully ergodic and extended, the density imbalance shows super-diffusive dynamics with $\gamma >1/2$ for all values of $\alpha$ studied. Corresponding imbalance plots are shown in \Fig{I1} for $h=t_0$ for various values of $\alpha$. $I(t)$ decays very fast during the initial time period and reaches around $0.01$ for $t_0t\sim O(10)$ after which it shows power-law decay with $\gamma > 1/2$ as shown in the inset of \Fig{I1}. For $h=t_0$, $\gamma\sim 0.68$ for all the values of $\alpha$ studied.  
 \\
 \\
  {\bf{Comparison of density imbalance for the random and uniform power-law interactions:}}\\
   \Fig{I2} shows the comparison of the density imbalance for the system with uniform power-law interactions and random power-law interactions for the same value of $h=5t_0$ and $\alpha=0.5$. In the case of uniform power-law interactions, the density imbalance saturates to a value close to unity after an initial decay. This shows that the system remains in the MBL phase having strong memory of the initial order. But in the presence of random power-law interactions, $I(t)$ shows a clear  power-law decay with the dynamical exponent $\gamma\sim 0.35$ indicating that the system is not in the MBL phase but shows sub-diffusive transport.


\begin{thebibliography}{99}
\bibitem{note}
It should be noted that though the Jordan-wigner transformation does not work for the long-range interacting systems, and hence the model of spin-1/2 particle studied in ~\cite{logan_lr} can not be mapped to the model of spin-less fermions studied in ~\cite{garg_lr}, qualitatively the same physics emerges in the two models as far as the effect of long-range interactions on MBL is concerned.
\end{thebibliography}
\end{document}